\documentclass[final,3p,times]{elsarticle}

\usepackage{amssymb}
\usepackage{braket}
\usepackage{multicol}

\begin{document}

\begin{frontmatter}

\title{Charged-particle detection efficiencies of close-packed CsI arrays}

\author{P. Morfouace$^1$, W. G. Lynch$^1$, M. B. Tsang$^1$}

\address{$^1$National Superconducting Cyclotron Laboratory and Department of Physics and Astronomy, Michigan State University, East Lansing, MI 48824, USA}

\begin{abstract}
Detector efficiency determination is essential to correct the measured yields and extract reliable cross sections of particles emitted in nuclear reactions. We investigate the efficiencies for measuring the full energies of light charged particle in arrays of CsI crystals employed in particle detection arrays such as HiRA, LASSA and MUST2.  We perform these simulations with a GEANT4 Monte Carlo transport code implemented in the NPTool framework. Both Coulomb multiple scattering and nuclear reactions within the crystal can significantly reduce the efficiency of detecting the full energy of high energy particles. The calculated efficiencies decrease exponentially as a function of the range of the particle and are quite similar for both the hydrogen ($p, d, t$) and helium ($^3$He, $\alpha$) isotopes. The use of a close-packed array introduces significant position dependent efficiency losses at the interior boundaries between crystals that need to be considered in the design of an array and in the efficiency corrections of measured energy spectra.
\end{abstract}

\begin{keyword}



\end{keyword}

\end{frontmatter}

\section{Introduction}
\begin{figure}[h!]
\begin{center}
\includegraphics[width=8.6cm]{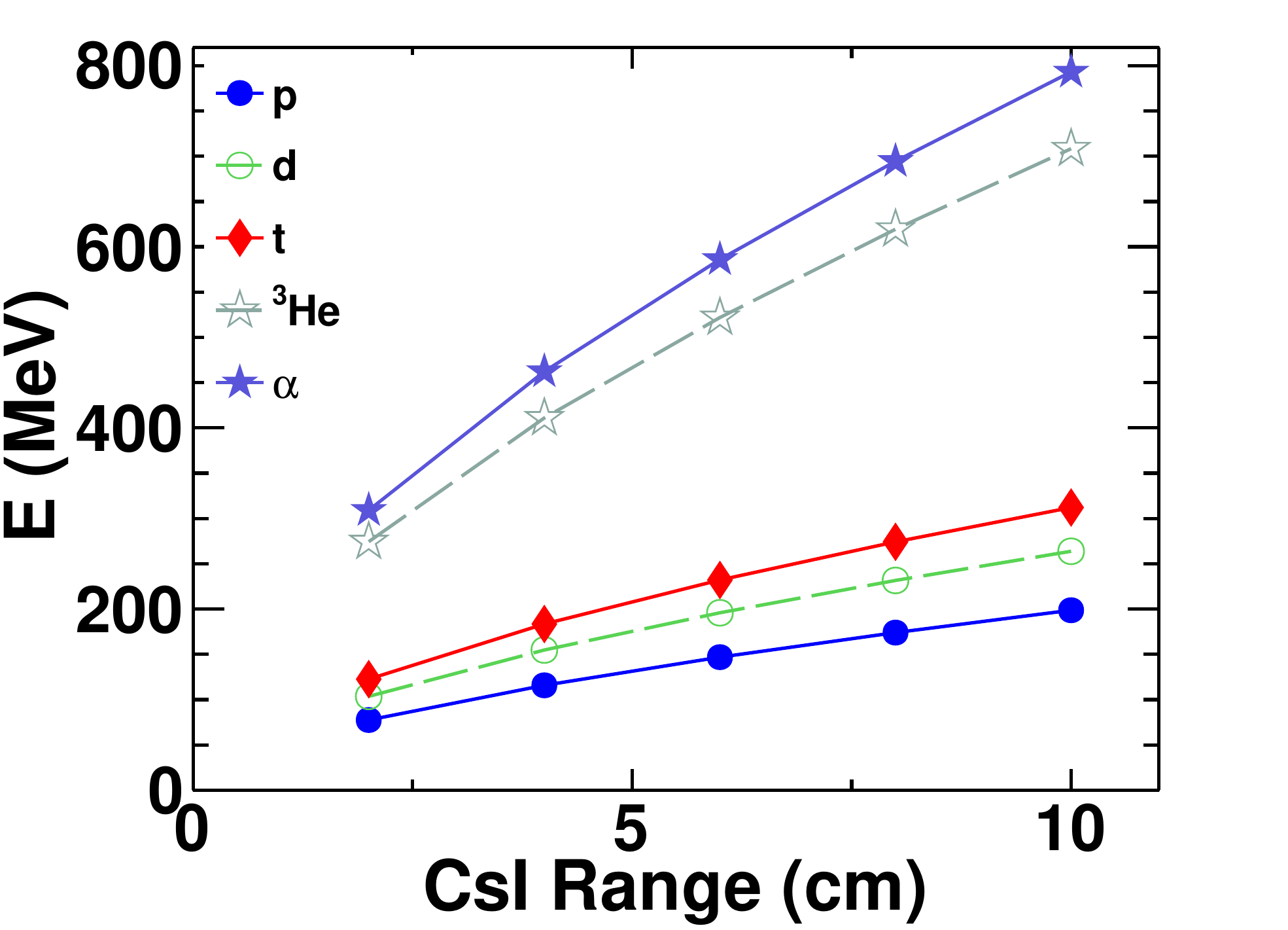}
\caption{\label{fig:CsIRange} Energy range for $p$ (blue line, closed circle), $d$ (dashed line, open circle), $t$ (red line, closed diamond), $^3$He (dashed line, open star) and $\alpha$ (solid line, closed star) as a function of the CsI thickness.}
\end{center}
\end{figure}

Silicon strip and pixel detectors are widely used to provide accurate position information regarding charged particles emitted in nuclear and particle physics experiments. To measure energetic light charged particles, such as protons or alphas with $E/A > 18$ MeV, which will penetrate through the thickest commercially available Si strip detectors, Si detectors are often backed by scintillators with thickness between 1 to 10 cm. Of the various scintillators suitable for detecting charged particles, Thallium doped CsI crystals have the virtue of scintillating at wavelengths that can be measured with silicon photodiodes. With energy resolutions for charged particle that are typically better than 1\%, they are easily machined and  are only mildly hygroscopic in air. Their cost also make them a popular choice to construct highly-efficient arrays \cite{lassa,hira07,must2}. Fig.~\ref{fig:CsIRange} shows the maximum measurable energies of $p, d, t, ^{3}$He and $\alpha$ particles as a function of the CsI thickness. As the electronic stopping power decreases inversely with the energy of the detected particle, the scintillator thickness required to stop a particle increases rapidly with energy. While longer crystals allow detection of higher energy particles, the efficiency for the measurement of the full energies of these particles decreases with energy due to scattering and reaction losses within the scintillator.

This paper focuses on calculating the loss in detection efficiencies of light particles as functions of the particle type and the detector thickness. As the energy of a particle increases, its range in the CsI crystal increases and the probability that the particle will undergo a nuclear reaction becomes more significant \cite{cha95}. In addition, there can be losses due to Coulomb multiple-scattering that occurs whenever the scattering deflects the charged particle out of the CsI crystal before it deposits its total energy. Such multiple scattering effects have been neglected in some previous calculations of reaction losses \cite{cha95} where it may have been justified by the geometry of the crystal. We show, however, that this can be important whenever the charged particles pass sufficiently close to the inner boundaries of the crystals in close-packed arrays.

\section{Simulation}
The method described in this paper can be applied to any Si-CsI detection system. To provide concrete examples, we have performed the simulations on individual telescopes of the Large Area Silicon Array (LASSA) \cite{lassa, lassa1}, the High Resolution Array \cite{hira07} and the MUr \`a STrip (MUST2) \cite{must2}. The configuration for each telescope used in this work consists of a Double Sided Silicon Detector (DSSD) backed by a close-packed array of 4, 4 and 16 CsI scintillator crystals corresponding to detection in the HiRA, LASSA or MUST2 arrays respectively. These telescopes were chosen because the authors are familiar with these devices and because they have geometries that are similar to other existing or planned arrays where we expect the present calculations can be of assistance in estimating the magnitude of such effects. We note however that most measurements with HiRA have utilized shorter 4 cm CsI crystals (HiRA). The recently completed upgraded HiRA array referred as HiRA10 in this paper has 10 cm crystals. Relevant details of the telescopes used in the simulations are listed in Table ~\ref{tab_telescope}.
 
\begin{figure}
\centering
\includegraphics[width=\textwidth]{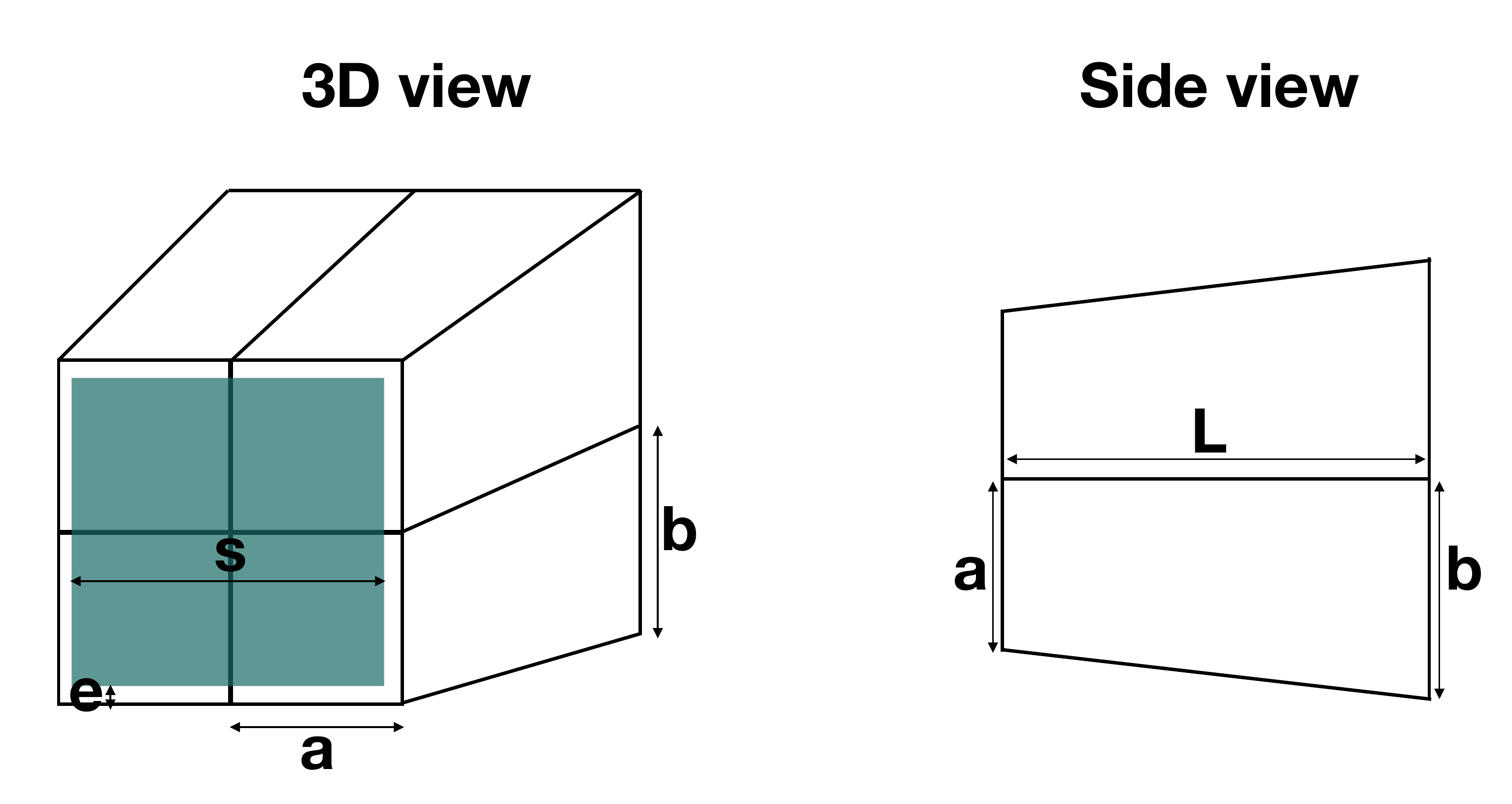}
\caption{\label{fig:geometry_crystal} Schematic drawing of a 2$\times$2 configuration telescopes displaying the relevant dimension parameters for the design of such an array.}
\end{figure}

A typical schematic drawing of a close-packed 2$\times$2 crystal configuration is shown in Fig.~\ref{fig:geometry_crystal} to illustrate the design of the HiRA and LASSA arrays. The square DSSD has an active area of $s \times s$ mm$^2$ backed with 4 identical crystals. Each crystal is $L$ mm long, with a front width of $a$ mm and a back width of $b$ mm. In order to minimize the multiple scattering effects on the outer edges of the crystal, the front width of two crystals is larger than the active area of the Silicon detector as illustrated in Fig.~\ref{fig:geometry_crystal} with $2e+s=2a$. The larger edges of the crystals allows the particles to go deeper into the crystal while reducing the efficiency loss near the outer edges of the crystals due to multiple scattering. Some details for the geometry of the HiRA10 and LASSA arrays are listed in Table.~\ref{tab_crystal_geo}.

For these calculations, we have adopted the NPTool framework \cite{npt16} that takes the full advantages of both ROOT analysis framework \cite{ROOT} and GEANT4 simulation toolkits \cite{Geant4}. The NPTool framework utilizes GEANT4 version 10.01, a Monte Carlo particle transport model that include the electromagnetic processes or hadronic processes or both. These processes can occur when the particles travel through the detector materials. In addition, the calculations for the MUST2 array utilize C++ classes that were developed within the NPTool framework \cite{npt16}. We developed analogous new classes for the LASSA and HiRA10 detectors so that all the different simulations and analysis are done consistently. These calculations take into account the known intrinsic energy resolution of different elements of the telescopes.

In this work we will focus on the detection of light charged particles with mass number $1<A<4$, i.e. protons to $\alpha$ particles. Similar calculations can be performed for any kind of charged particle. In the following, the effects of Coulomb multiple scattering are explored separately in Section 2.1 and in conjunction with reaction losses in section 2.2.

\begin{table*}[t!]
\centering
\begin{tabular}{c | c | c | c}        
 Detector					& DSSD  			& CsI  			& Design distance  \\  
 						& thickness $\mu$m	& thickness (cm)	& from the target \\
						&				&				& to the silicon detector (cm) \\ \hline \hline
MUST2 \cite{must2}			& 300		& 4	& $\approx 17$ \\ 
LASSA \cite{lassa}			& 500		& 6	& $\approx 20$ \\ 
HiRA10 \cite{hira07}			& 1500		& 10	& $\approx 35$ \\ 
\end{tabular}
\caption{\label{tab_telescope} Relevant information for various telescopes.}
\end{table*}

\begin{table*}[t!]
\centering
\begin{tabular}{c | c | c | c | c | c}        
 Detector			& $a$ (mm)  		& $b$ (mm)  	& $s$ (mm)		& L (mm)   	& $e$ (mm)	\\  \hline \hline
LASSA 			& 26.5			& 33.8		& 50				& 60			& 1.5				\\ 
HiRA10 			& 34.9			& 44.6		& 64				& 100		& 2.9				\\
\end{tabular}
\caption{\label{tab_crystal_geo} Details of the geometry of the crystals. See text and Fig.~\ref{fig:geometry_crystal} for parameter definition.}
\end{table*}

\begin{figure}
\centering
\includegraphics[width=\textwidth]{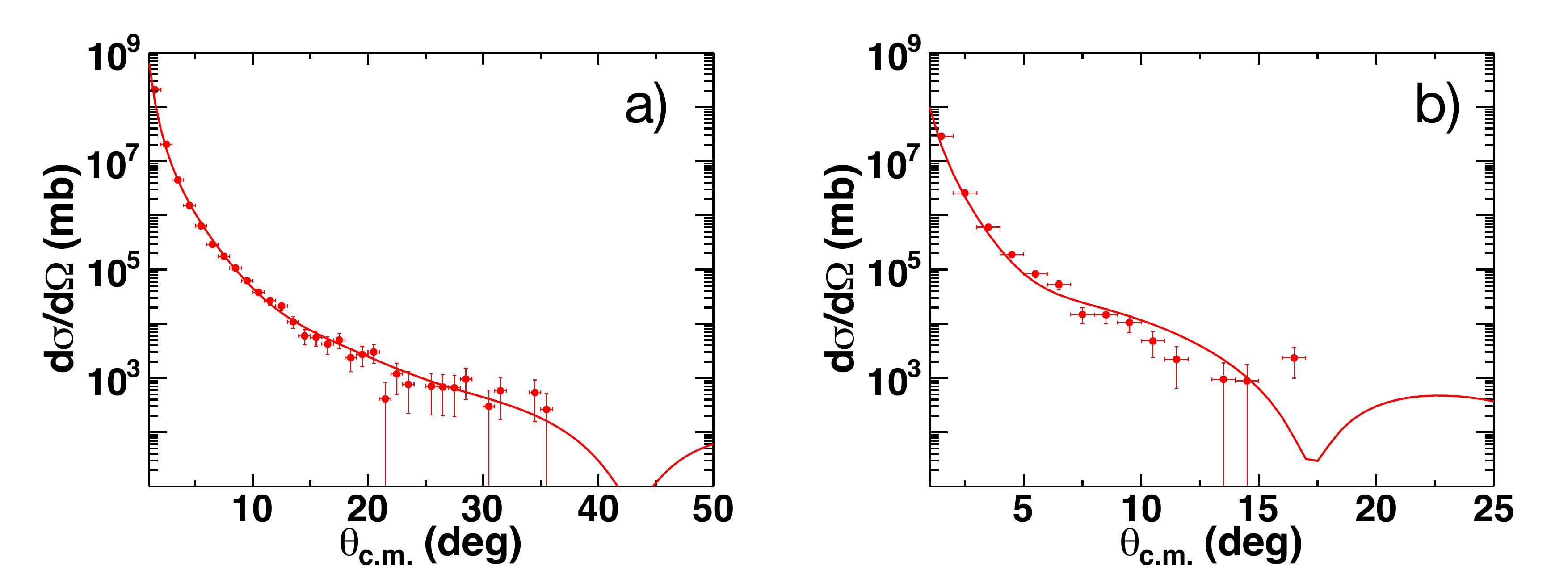}
\caption{\label{fig:scattering} (a) Angular cross section of the scattered proton for the $^{133}$Cs target at 30 MeV and (b) for the $^{133}$Cs target at 80 MeV proton incident energy. In both figures the points correspond to the GEANT4 simulations while the solid lines correspond to the DWUCK calculations.}
\end{figure}

\subsection{Multiple-Scattering}
In this subsection, we focus on the multiple-scattering that a particle experiences when going through a given material. The single and multiple Coulomb scattering influences more strongly lighter charged particles such as protons that have smaller momenta.  To simulate the process we have used the standard electromagnetic package {``option4''} in GEANT4. In this package the finite size of the nuclei in the detector material and the penetration of the detected particle through the Coulomb barrier of such a nucleus is modeled by the Born approximation. The package sets the nuclear form factor to zero when the scattering angle of the particle from that nucleus is greater than the angle $\theta_{max}$, which is defined to be (chapter 6 of ref.~\cite{geant4_doc}):
\begin{equation}
\sin\big(\frac{\theta_{max}}{2}\big)=\frac{1}{kR},
\end{equation}
where $k$ is the wave number of the incident particle and $R$ is the radius of the target nucleus. Within GEANT4, this form factor modification replaces an optical model calculation of scattering from the Coulomb and nuclear optical potential, whose imaginary term models the loss of flux into other channels. It is therefore important to check that this approximation does not overestimate the elastic scattering at large angles where the optical potential strongly modifies the elastic scattering cross section.

To assess the accuracy of this approximation, we simulated the scattering of both 30 MeV and 80 MeV proton particles in a 5 $\mu$m thick $^{133}$Cs target. We compare the results from our GEANT4 simulations to optical-model calculations in Fig.~\ref{fig:scattering}. Here, the GEANT4 simulations for $10^7$ incident particles are shown by the points with statistical error bars. Corresponding optical model calculations, shown by the solid lines, are performed with the elastic scattering predictions of the DWUCK code \cite{dw4} using the CH89 global optical potential \cite{ch89}, which is well adapted for the energy and the mass region we are considering. While the GEANT4 yields do not have diffraction minima that match the minima in the calculated differential cross sections, the average trend with angle is rather well matched suggesting that the GEANT4 simulation provides a reasonable approximation of the scattering of particles to large angles.

\begin{figure}
\begin{center}
\includegraphics[width=\textwidth]{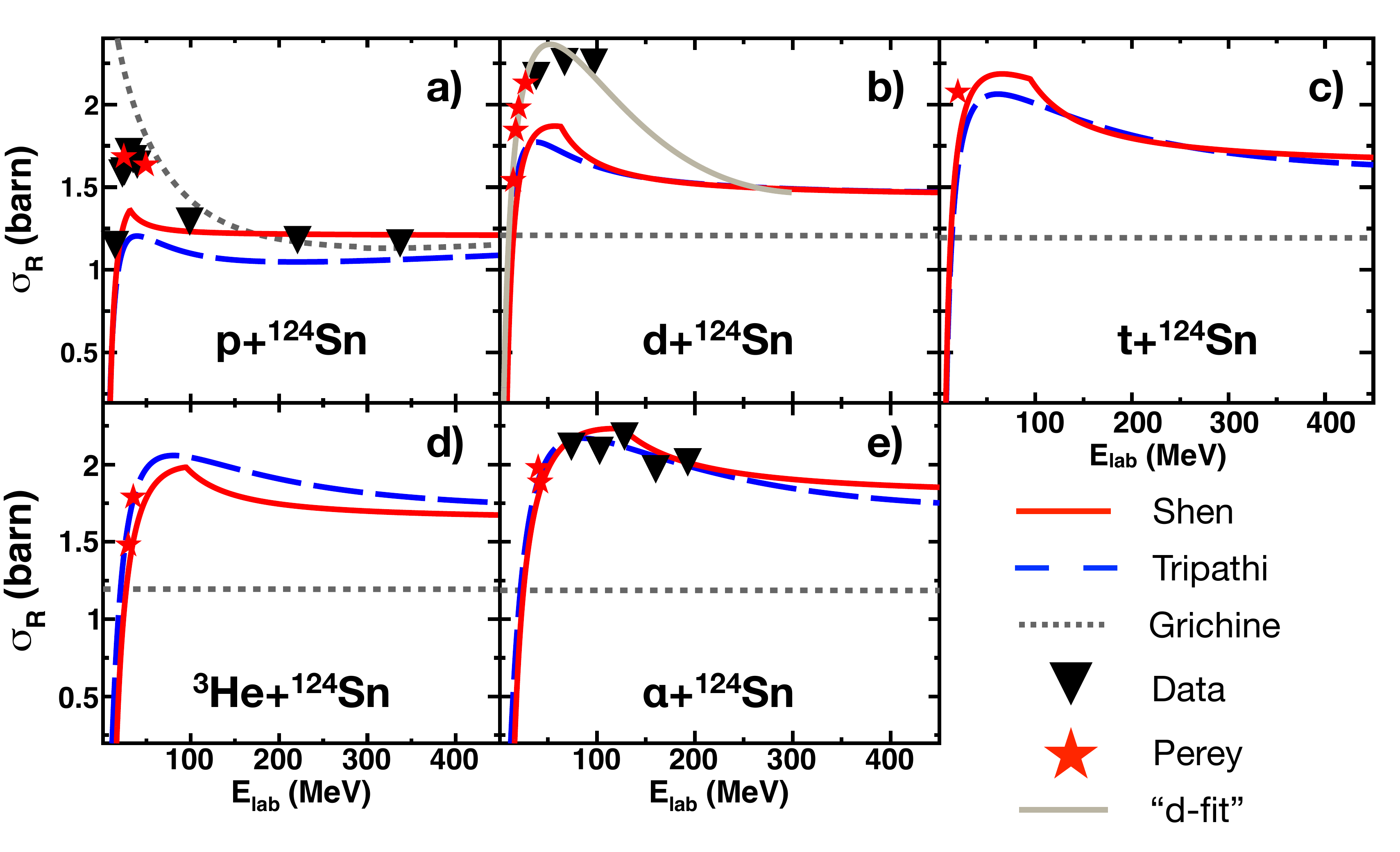}
\caption{\label{fig:G4Crossection} Reaction cross sections for (a) $p$, (b) $d$, (c) $t$, (d) $^{3}$He and (e) $\alpha$ beams on $^{124}$Sn target as a function of the beam energy using different parametrization in GEANT4. The black inverted triangles correspond to experimental data from Ref.~\cite{carlson} for protons, Ref.~\cite{madani} for deuterons and Ref.~\cite{tripathi} for $\alpha$ particles. The star symbols correspond to the total cross section calculated using Perey and Perey optical parameters from Ref.~\cite{per76}.}
\end{center}
\end{figure}

\begin{figure}
\begin{center}
\includegraphics[width=\textwidth]{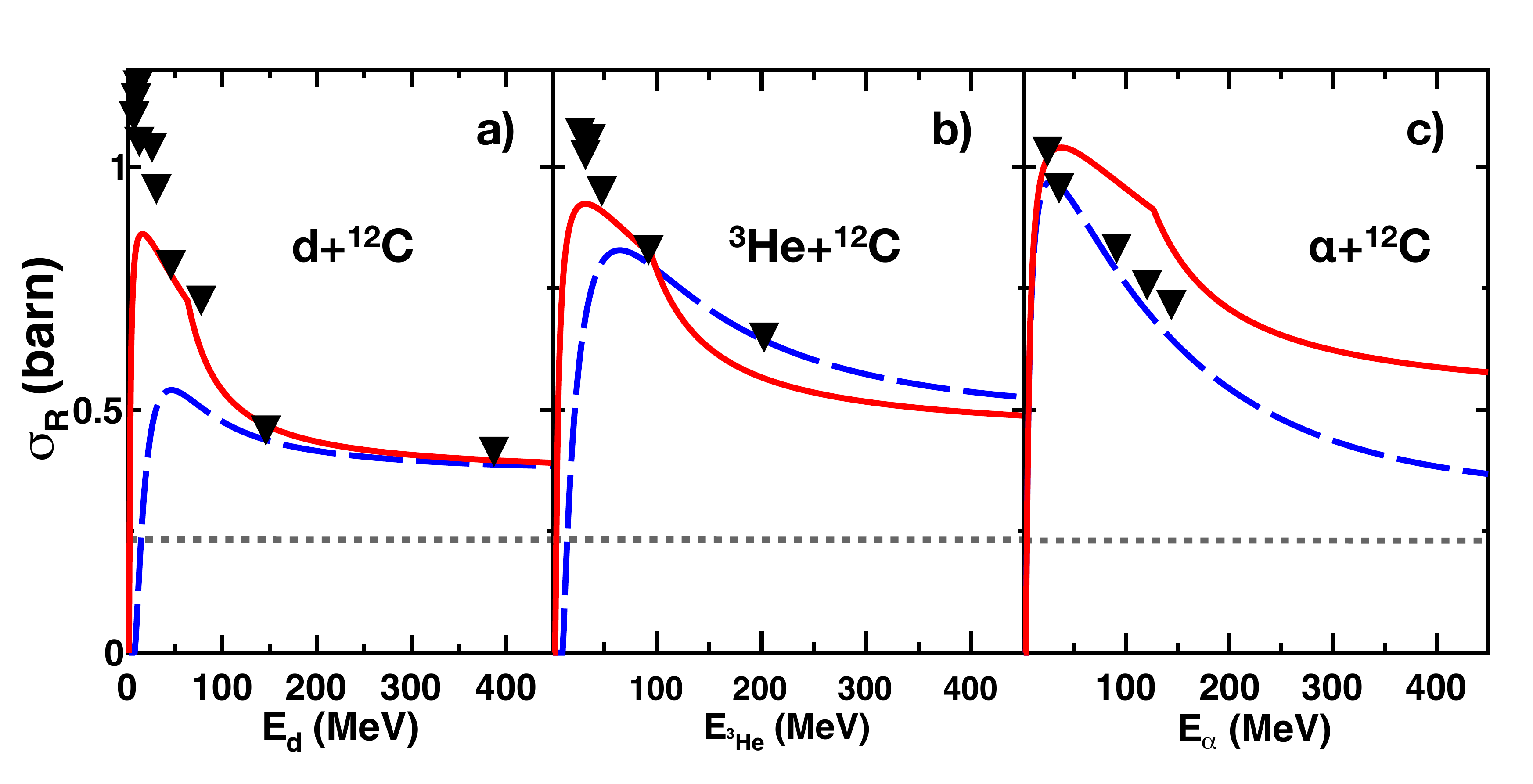}
\caption{\label{fig:G4Crossection_12C} Reaction cross sections for (a) $d$, (b) $^{3}$He and (c) $\alpha$ on $^{12}C$ as a function of the particle energy using different parametrization in GEANT4. The inverted triangles correspond to experimental data coming from Ref.~\cite{pen81}.}
\end{center}
\end{figure}

\subsection{Nuclear reactions}
\label{sec:nuclear_reaction}
In this subsection, we explain how the reaction losses are estimated. We compare the different parameterizations available in GEANT4 to experimental data, when available,  in order to make a suitable choice for each of the particles we are considering and to estimate the uncertainty of the efficiency calculations. The G4BinaryLightIonReactionModel option in GEANT4 has been used to evaluate the predicted efficiency losses for different cross section parameterizations. Two parameterizations are from Shen \cite{shen}, and Tripathi \cite{tripathi}, which provide estimates of the reaction cross sections at energies ranging from few AMeV to few AGeV. Both models apply a correction to reduce the reaction cross sections at low incident energies near the Coulomb barrier. At much higher energies, the cross section is reduced again, reflecting the energy dependence of the average nucleon-nucleon cross section \cite{devries}. To describe protons, we have also used a  third parametrization by Grichine \cite{glauber}, which is based on a simplified Glauber approximation.

We are mainly interested in the reaction cross section of light charged particles with CsI. However since there are no suitable experimental reaction cross sections for light charged particles with either Cs or I, we compare the GEANT4 prediction to $^{124}$Sn target, for which some experimental light particle data exist. As $^{124}$Sn has a mass similar to $^{137}$Cs, the cross section prediction for Cs is very similar to the one for Sn. However, the difference in $Z$ between Sn and Cs increases the predicted cross sections by about 10\% relative to Sn. Fig.~\ref{fig:G4Crossection} shows the predicted cross sections of these three models (Shen, Tripathi and Grichine) for hydrogen and helium beams incident on a Tin target, as a function of incident beam energy. The solid inverted triangles show the experimental data for protons \cite{carlson},  deuterons \cite{madani}, and $\alpha$ particles \cite{tripathi}. Except for proton and $^3$He projectiles, cross sections provided by the Shen \cite{shen} and Tripathi \cite{tripathi} parameterizations are similar. Unlike the others, the Grichine parameterization \cite{glauber}, based on Glauber model, is constant for $d, t, ^{3}$He and $\alpha$ particles. Only the proton cross section displays a significant energy dependence. The predicted Grichine cross section for protons is comparable to the proton data, but its predictions are low and unrealistic for the other charged particles. The Grichine prediction will therefore only be used for the efficiency calculation involving protons. We also tested the intra-nuclear cascade (INCL) model using G4HadronPhysicsINCLXX and G4IonINCLXXPhysics \cite{incl} and found that the results are similar to that of Shen and Tripathi.

Since the experimental total cross sections on tin are limited, we also compare in Fig.~\ref{fig:G4Crossection_12C} the total cross sections for $d$, $^3$He and $\alpha$ beams on $^{12}$C from \cite{pen81}. We did not find suitable experimental data for tritons on $^{12}$C. Once again the predicted cross sections for $d, ^{3}$He and $\alpha$ particles provided by the Grichine option are unrealistic. Here, the Tripathi parameterization provides a better agreement with the $\alpha$ particle data. The experimental deuteron and $^3$He data are better described at low energies by Shen parameterization than by the Tripathi one, but at higher energies the Tripathi parameterization provides a somewhat better description of the $^3$He and $\alpha$ particle data. At low energies ($<50$ MeV) the Shen cross section underestimates the measured deuteron total cross section by about 20\% and the disagreement with the Tripathi parameterization is even larger. In the next subsection, however, we will show that the contribution from the different reaction cross section parameterizations to the efficiency from this energy range are insignificant (cf Fig.~\ref{fig:Eff_H}).   

\section{Results}
Unless otherwise indicated as in the case of calculations using Charity parameterization, one should assume that Coulomb multiple scattering effects are included in calculations shown in this section.

\subsection{Effects on particle identifications}

\begin{figure}
\begin{center}
\includegraphics[width=\textwidth]{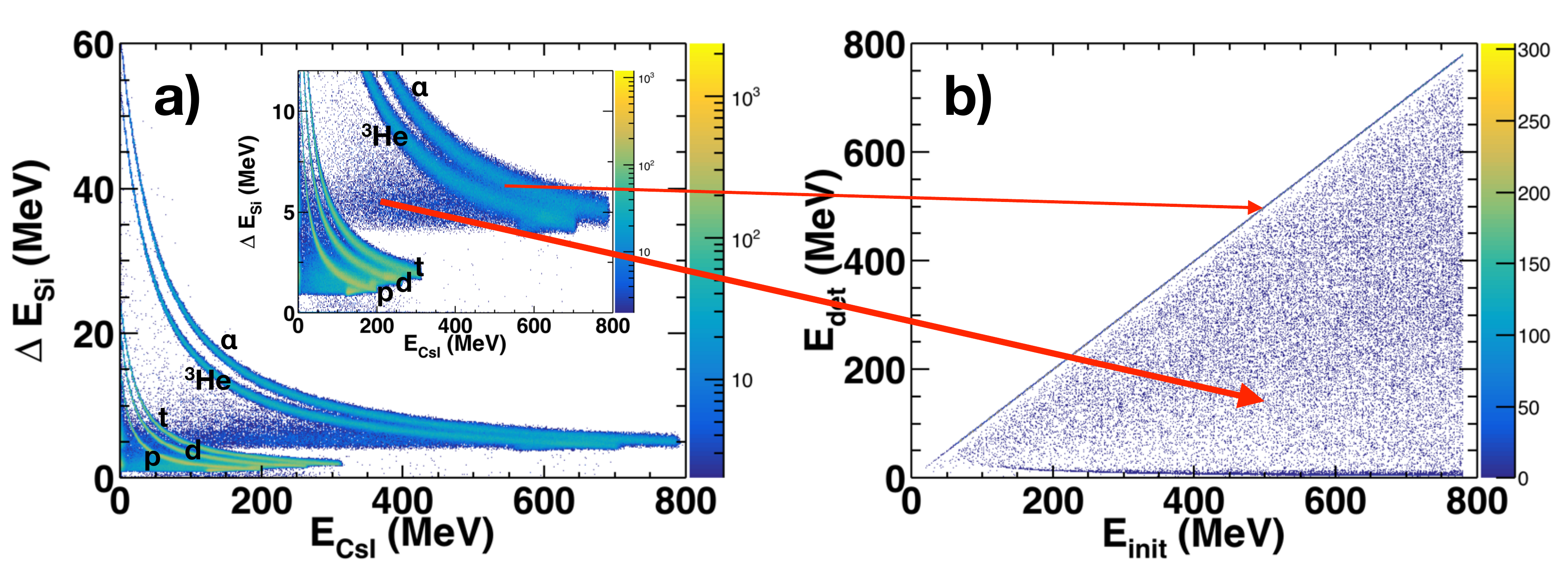}
\caption{\label{fig:hira_pid} (a) Particle identification spectrum from the $\Delta E-E$ method with a zoom as an inset. (b) Correlation between the initial simulated energy versus the reconstructed energy. See text for details.}
\end{center}
\end{figure}

To calculate the efficiency for detection of the full energy of the particle, we simulate the interaction of the various particle species ($p, d, t, ^{3}$He or $\alpha$ ) in CsI by assuming an initial flat energy distribution between $0 <  E_i <E_{max}$ MeV, where $E_{max}$  is the energy for which the range in CsI equals the thickness of the CsI crystal. Particles are assumed to be emitted from a target located at the design distances of 35 cm, 20 cm and 17 cm in front of the HiRA, LASSA and MUST2 telescopes, respectively. We reconstruct the total kinetic energy of each particle by adding the calculated energy lost in the DSSD Si detector ($\Delta E$) and the energy detected in the CsI crystals ($E$). 

The various particle species can be distinguished via a Particle Identification Spectrum (PID) constructed by plotting calculated values in a two-dimensional $\Delta E$ vs. $E$ spectrum. Such a spectrum is shown in Fig.~\ref{fig:hira_pid} for simulations involving the HiRA10 telescope. The inset in the figure depicts an expanded view at low values of $\Delta E$ that allow one to view the energy loss of energetic particles that stop completely in the CsI crystal. Clearly, one can easily distinguish the particle species. 

The blue haze outside the particle lines corresponds to events when a nuclear reaction occurs in the crystal affecting the measured energy or when the particle scatters out of the side of the crystal. Both effects result in the mis-identification of the particle. This incomplete energy collection is clearly illustrated in the right panel of Fig.~\ref{fig:hira_pid}, where the initial energy $E_i$ is plotted against the reconstructed energy $E_{det}$ . The 45$^{\circ}$  line corresponds to the well reconstructed events with good PID while all the events below the line correspond to the blue haze in the PID plot where the energy is not reconstructed correctly. In the following we label a particle as fully detected when $|E_{det} - E_i | < 2$ MeV, where $E_{det}  = E_{Si} + E_{CsI}$. This range is wide enough to prevent particles on tails of resolution functions from being mislabeled as an out-scattered or reaction event.

\subsection{Efficiency as a function of energy}

\begin{figure}
\includegraphics[width=\textwidth]{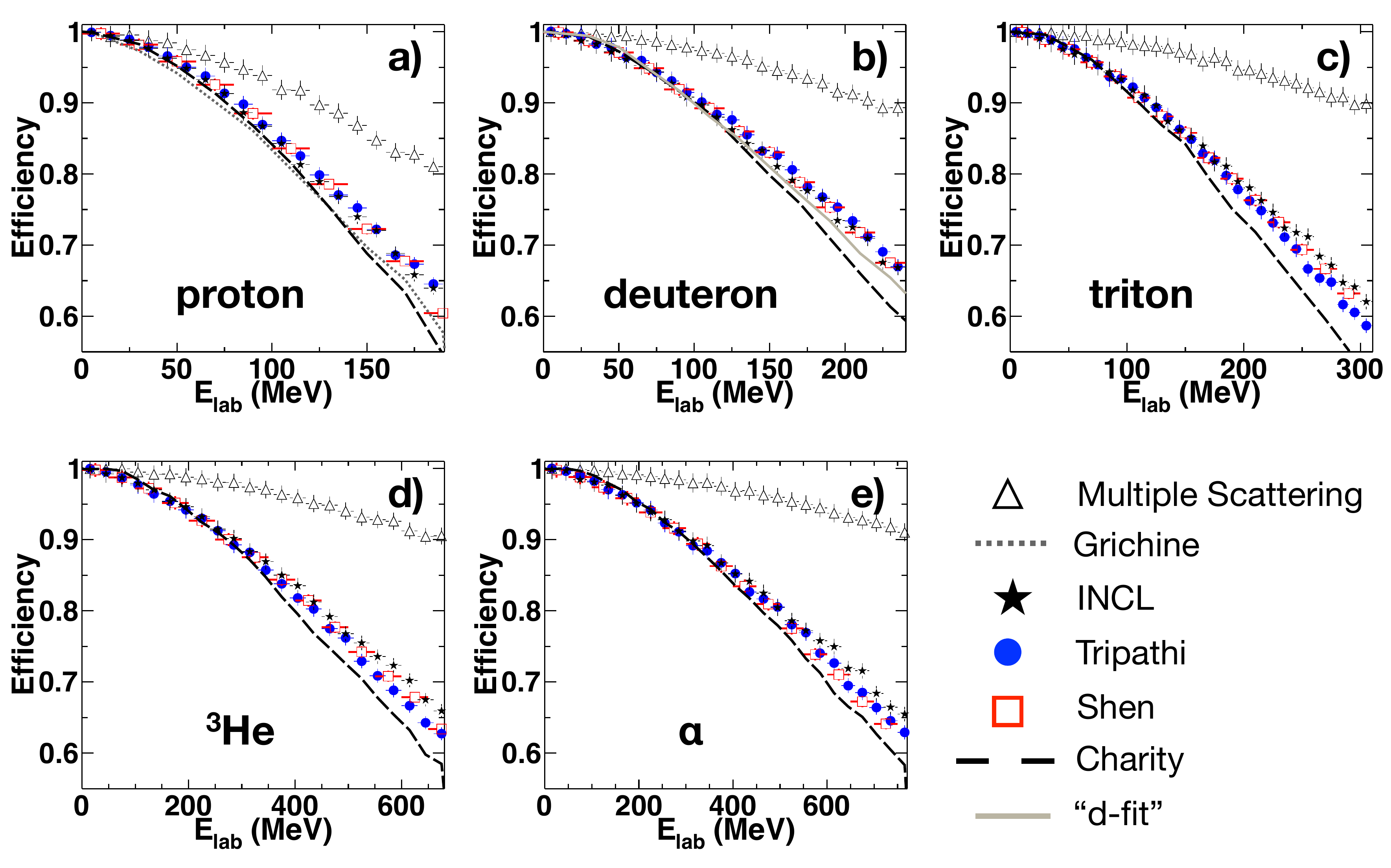}
\caption{\label{fig:Eff_H} Efficiencies to detect the full energy of the proton (a), deuteron (b), triton (c), $^{3}$He (d) and $\alpha$ particles (e) in the HiRA10 as a function of the energy. The open triangle symbols show the efficiency when only the multiple scattering (MSC) is taken into account. The full blue circles, open red squares and black full stars correspond to the efficiency using the Tripathi \cite{tripathi}, Shen \cite{shen} and INCL parameterization respectively. The black dashed lines correspond to the determination of the efficiency using the Charity parameterization. For protons, the dotted dashed line corresponds to the efficiency using the Grishine \cite{glauber} parameterization. For deuterons, the light grey line corresponds to the result using the {``d-fit''} parameterization of the experimental deuteron cross section shown in Fig.~\ref{fig:G4Crossection}.}
\end{figure}

Fig.~\ref{fig:Eff_H} shows the efficiencies for well detected $p, d, t, ^{3}$He and $\alpha$ particles in the HiRA10 crystals up to maximum energies of 198, 263, 312, 708 and 793 MeV, respectively. These efficiencies are also valid for the original HiRA CsI crystals up to the maximum energies of 115, 155, 183, 410 and 462 MeV for $p, d, t, ^{3}$He and $\alpha$ particles, respectively, corresponding to a range of 4 cm in CsI. Clearly, the calculated detection efficiencies for all particle species decrease with incident energy reflecting an increased average number of interactions as the particles penetrate further into the crystals.

The open triangles in each panel show the efficiency losses that are solely due to the effects of multiple scattering out of the crystal. Protons have significantly larger efficiency losses due to multiple scattering than do the other particle species. This reflects the smaller momenta of protons for a given kinetic energy, which are more comparable to the probable momentum transfers resulting from Coulomb interactions with the Cs and I atoms in the CsI crystals.

The other symbols show the lower efficiencies that result from hadronic reaction losses using the Tripathi (solid circles), Shen (open squares), INCL (solid stars) and Grichine (dotted line) parameterizations for the nuclear cross section \cite{tripathi, shen, glauber}. We note that the results for the Grichine parameterization are only calculated for protons because this parameterization severely under-predicts the reaction cross sections of the other species.

In Ref \cite{cha95}, Charity \textit{et. al} calculated efficiency losses due to nuclear reactions using optical model parameterizations of Perey and Perey \cite{per76}. We parameterize the nuclear reaction losses of Charity \cite{cha95} and incorporate them into the GEANT4 simulation including multiple scattering. These efficiencies are shown as dashed lines in Fig.~\ref{fig:Eff_H}. Details of the implementation of the Charity parameterizations in this work are discussed in \ref{appendixA}.

All the particle efficiencies derived from Shen, Tripahti and INCL agree to within 1\%.  This is consistent with Fig.~\ref{fig:G4Crossection}. For protons, the Grichine parameterization provides the lowest efficiency. This can be expected from Fig.~\ref{fig:G4Crossection}a), where it over-estimates the measured reaction cross section at low energies. Except in the case of the proton efficiencies and the efficiencies of other species at low energies, the Charity parametrization provides the lowest efficiency values. This is not surprising because the cross section is assumed to remain high and constant  as the energy is increased, which over estimates the reaction cross section at higher energies (cf Tab.~\ref{tab:charity} for Charity cross-section value).

\begin{figure}
\begin{center}
\includegraphics[width=8cm]{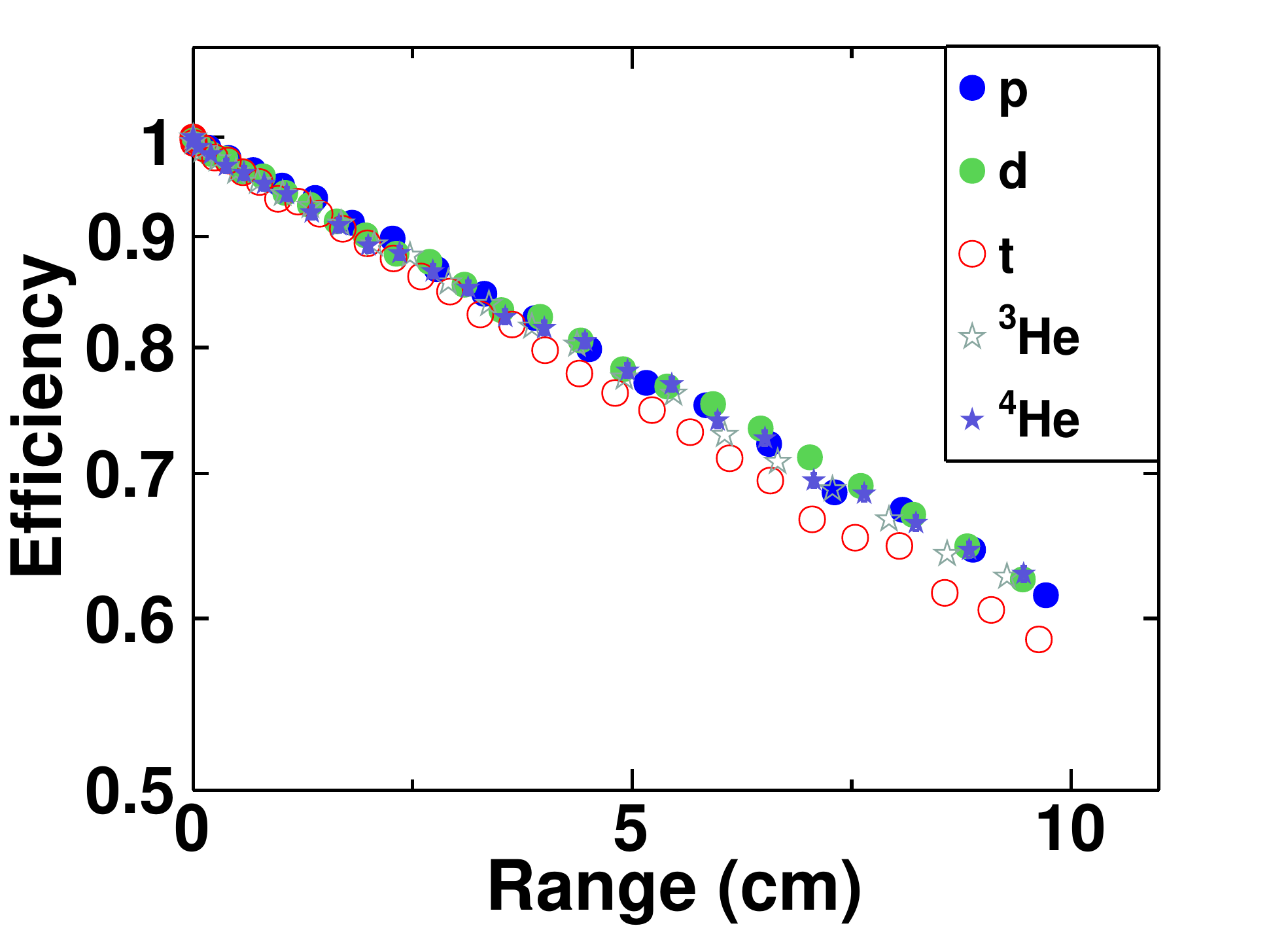}
\caption{\label{fig:Eff_Range} Efficiency to detect the full energy of $p, d, t, ^{3}$He and $\alpha$ particle in HiRA10 as a function of the range in cm in the CsI crystal using the Tripathi parametrization \cite{tripathi}.}
\end{center}
\end{figure}

In the case of deuterons, no parameterizations describe the existing reaction cross section data. In order to estimate the change that could be expected for a cross section that reproduces the reaction cross section at low energies, yet decreases with energy at higher energies, we have modeled the cross section for deuterons by the grey solid line shown in the deuteron panel in Fig.~\ref{fig:G4Crossection} and labeled as {``d-fit''}. With this parameterization, we followed the technique used for Charity parameterization detailed in \ref{appendixA}. As shown in Fig.~\ref{fig:Eff_H} for deuterons, the calculated efficiency using the {``d-fit''} cross section gives somewhat lower efficiencies than that of Tripathi and Shen, but the difference is less than 2\%. This comparison shows the consistency of the calculations using different parameterizations and suggests that the efficiency can be determined to within a few percent at the highest energies.

We regard the {``d-fit''} parameterization as a reasonable best estimate of the efficiency for deuterons. For protons, the best estimate lies midway between the Shen and Grichine parameterizations while for the $t, ^{3}$He and $\alpha$ particles, our best estimate for the efficiency lies between the calculations for the Shen and Tripathi parameterizations. Based on the variance of the efficiency calculations, we estimate the uncertainty in the efficiency to be $\delta = 2.5\%\frac{E}{E_{max}}$.

Since the efficiency is reduced by scattering from the Cs and I atoms in the crystal, the efficiency should decrease exponentially as a function of the range in the CsI. This is shown  for the Tripathi parameterization in Fig.~\ref{fig:Eff_Range} for HiRA10 configuration. Interestingly, the efficiency is very similar for all the particles when plotted as a function of the range. As expected, the efficiency decreases exponentially with range from a value of 100\% at low energies to about 60\% at 10 cm. This general exponential decrease illustrates that the detailed dependence of the reaction cross section does not strongly influence the energy dependence of the efficiency.

\begin{figure}
\begin{center}
\includegraphics[width=8cm]{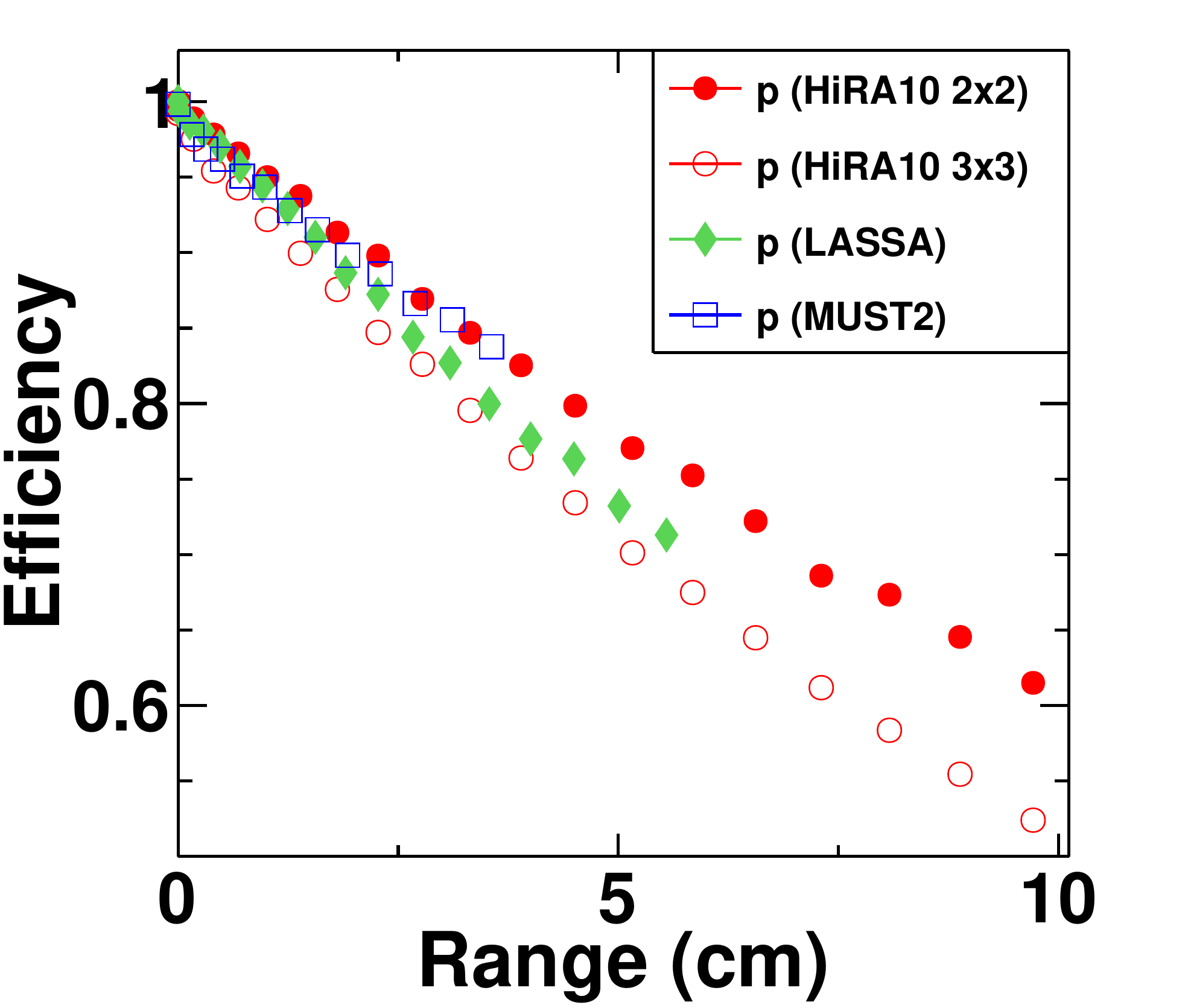}
\caption{\label{fig:Eff_telescopes} Efficiency to detect the full energy of a proton for the HiRA10 2$\times$2 (closed red circles), the HiRA10 3$\times$3 (open red circles), LASSA (green diamonds) and MUST2 (open blue rectangles) telescopes.}
\end{center}
\end{figure}

\subsection{Contributions from different reaction processes}
The incoming light charged particle can undergo an inelastic reaction, which is a part of the total reaction cross section. When a nucleus is excited below the particle emission threshold, it usually de-excites by $\gamma$ decays. The emitted $\gamma$ ray can be absorbed by the crystal. Its energy, or part of it, may then be included in the measured particle energy depending on where in the crystal that the inelastic scattering occurs and on the photo-peak efficiency of the crystal for $\gamma$ rays at that location. With our cut of $\pm 2$ MeV, some of these processes are included in the full energy peak. While the detailed nuclear structure of the Cs and I nuclei is not fully modeled in the simulation, some aspects of the $\gamma$ de-excitation process is handled through GEANT4 with the \textit{G4DecayPhysics} and \textit{G4RadioactiveDecayPhysics} classes.

For an inelastic collision which emits particles, the particle ID for the incident particle will be incorrect and will not be accounted for in our efficiency determination. However, the typical cross section for inelastic scattering leaving the target nucleus in low-lying states is about few millibarns \cite{ter73}, whereas the total reaction cross section is of the order of a barn. The neglect of a detailed accounting for inelastic processes can be shown to be a small effect by making a comparison to the larger uncertainties in the correction due to the reaction cross section. In the case of the deuteron induced reactions discussed in section 3.2, the total cross section changes by 25\% with the {``d-fit''} parameterization compared to Shen parameterization. However, the efficiency determination changes by less than 2\% at high kinetic energies. Keeping this in mind, and considering the small contribution ($\approx 1$\%) of the inelastic reactions to the total cross-sections, we expect the errors introduced by neglecting the detailed contribution of the inelastic processes to be small in comparison to the 2\% estimated uncertainty in the efficiency determination.

\subsection{Influence of granularity on the detection efficiencies}
There are many considerations in designing the scintillation array behind the Si strip detector. In heavy ion collision experiments and multi-particle resonance decay spectroscopy where charged particle multiplicities are high, it is desirable to construct an array with high granularity to minimize multiple hits in one crystal. In principle, such multiple hits render the data in a crystal invalid, and represent an addition source of efficiency loss beyond that discussed above. In all the telescopes we discussed here, the CsI crystals placed behind the Si detectors form a closed packed geometry; 2$\times$2 for LASSA and HiRA10, 4$\times$4 for MUST2, with 6, 10 and 4 cm long CsI crystals respectively. Ideally, the crystals should take the shape of a tapered cone forming part of a sphere with the target position at the design distance. In practice, each crystal has a trapezoidal shape. In the close-packed 2$\times$2 geometry of both LASSA and HiRA10, the inner surfaces between crystals are at right angles to each other and to the front and back surfaces as shown in Fig.~\ref{fig:geometry_crystal}. The outside surfaces flair out trapezoidally so as to contain the particles that pass through the active area at the front surfaces of the telescopes. To illustrate the dependence of the efficiency on the actual detector geometry, we plotted in Fig.~\ref{fig:Eff_telescopes} the efficiency of protons as a function of the range using Tripathi parameterization for HiRA10 2$\times$2 crystals (closed red circles), for a hypothetical HiRA10 3$\times$3 crystals (open red circles), LASSA (green diamond) and MUST2 (blue open square). Even though the layout and the construction of the HiRA and LASSA crystals are similar, there are differences in their efficiencies,  that are most evident for high energy protons. This reflects the size of the silicon detector and, in turn, the size of the crystals. Four individual LASSA crystals sit behind a single 50 mm $\times$ 50  mm silicon detector.  Consequently, their front faces are smaller at 26.5$\times$26.5 mm$^2$ than are the HiRA10 crystals, which measure 34.9$\times$34.9 mm$^2$ on the front surface behind the active area of a 64 mm $\times$ 64 mm silicon detector. We have modeled the efficiency of a hypothetical HiRA10 3$\times$3 array configuration. This latter configuration covers the solid angle with 9 crystals, each having smaller  22 mm $\times$ 22 mm front surfaces. These smaller crystals detect  energetic protons with a significantly lower detection efficiency than the HiRA10 2$\times$2 array configuration as illustrated in Fig.~\ref{fig:Eff_telescopes}.

The choice of crystal sizes and shapes reflects a compromise between granularity and detector size. One would like to maximize the efficiency with regard to budgetary and size constraints. On the other hand, one would like to have crystals with smaller individual solid angles to minimize coincidence summing, \textit{ie.} the probability that multiple particles hit the same crystal. This can lead to efficiency problems when particles scatter out of the crystals.

\begin{figure}
\begin{center}
\includegraphics[width=16cm]{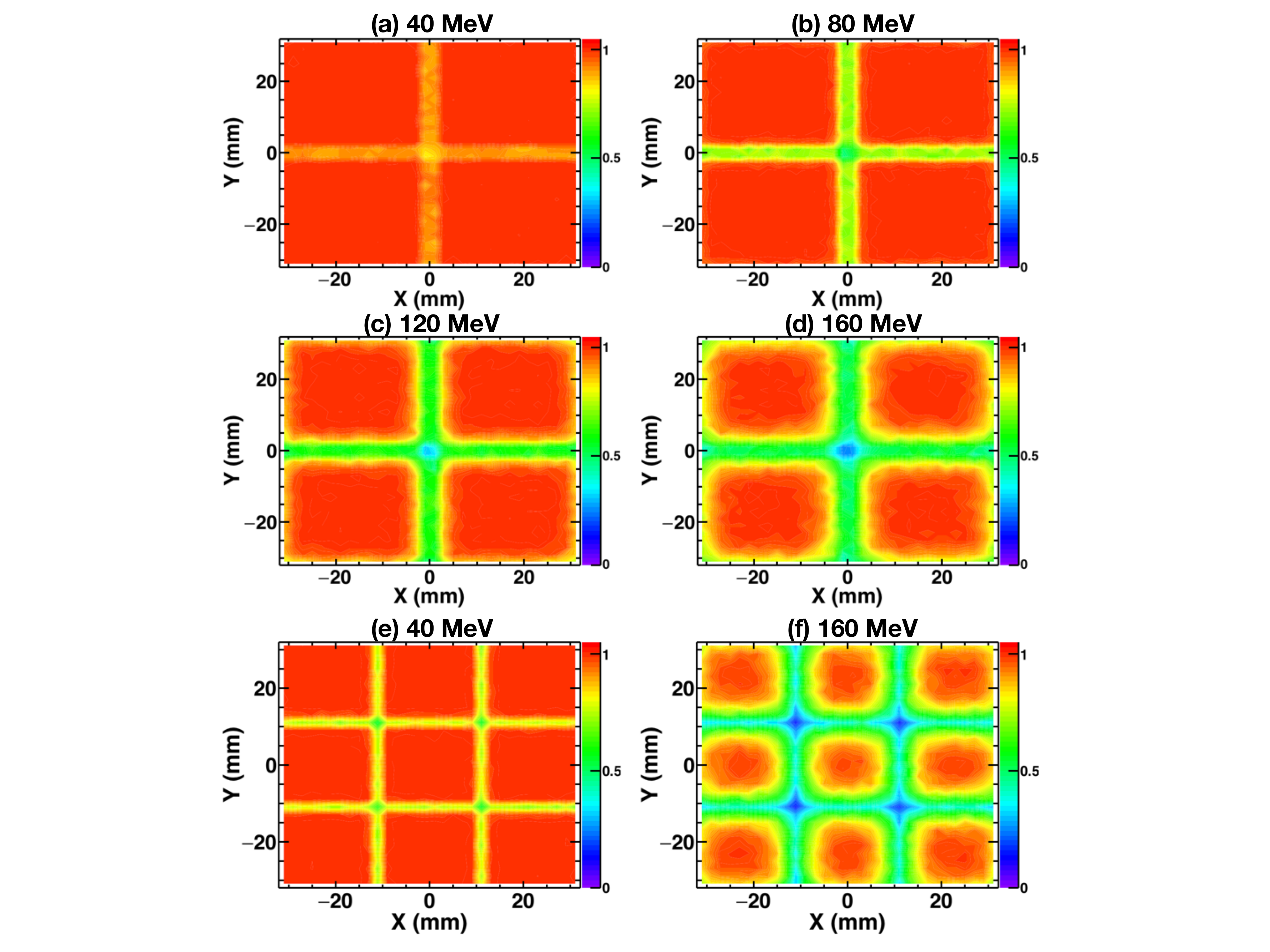}
\caption{\label{fig:Edge_effects} Front view of one HiRA10 telescopes showing the efficiency to detect the full energy for different proton energies at 40, 80, 120 and 160 MeV for a 2$\times$2 crystal configuration and at 40 and 160 MeV for a 3$\times$3 crystal configuration. See text for details.}
\end{center}
\end{figure}

To illustrate this tradeoff, we simulate the efficiency for detection of mono-energetic protons at 40, 80, 120 and 160 MeV with one HiRA10 telescope using a 2$\times$2 configuration (Fig.~\ref{fig:Edge_effects} (a-d)). We also simulate protons at 40 and 160 MeV using a 3$\times$3 crystals (Fig.~\ref{fig:Edge_effects} (e-f)). We show the efficiency as a function of the position where the proton hits the 6.4 cm $\times$ 6.4 cm HiRA silicon detector located 1 cm in front of the CsI crystals.

For the case of 40 MeV protons, the efficiency is uniform at nearly 100\% for most of the surface of the telescope, but it decreases to about 90\% near the inner boundaries between crystals reflecting multiple scattering out of the inner crystal boundaries. A comparable reduction of efficiency is not observed at the outer boundaries of the crystals because outer edge of the crystal is more than 2 mm outside of the active area defined by the passage of the charged particles through the silicon detector as explained in Fig.~\ref{fig:geometry_crystal} and the extra size grows to 12.6 mm at the far end of crystal. With increasing proton energies, the efficiency decreases. The decrease is more significant at the inner boundaries between crystals with the region of reduced efficiency grows in width, becoming about 5 mm wide and the efficiency decreases to about 50\% at 160 MeV

To minimize the rejection of the events with multiple hits, one can increase the granularity of the array. This means using more crystals to subdivide the active Si detector area more finely, minimizing possible multiple hits by detecting the particles emitted in close proximity in separate detectors. To illustrate such effects of increased granularity on efficiency, we consider a 3$\times$3 configuration shown in Fig.~\ref{fig:Edge_effects}(e-f) which has the same thickness and solid angle coverage as the HiRA10 array. The efficiency of such a 3$\times$3 array is shown in the bottom two panels of Fig.~\ref{fig:Edge_effects}. The decrease in efficiency near the inner boundaries of the 3$\times$3 array leads to a reduced efficiency near the edges of the crystal that is rather similar in width as for the 2$\times$2 configuration. However, the fraction of the area that lies near these boundaries is larger and the degradation in the average efficiency with energy much worse than for the 2$\times$2 configuration. Clearly, the loss of efficiency due to multiple scattering depends strongly on the ratio of the depth that the detected particle penetrates into the crystal divided by the width of the crystals. One way to improve granularity without worsening the out-scattering inefficiency is to build larger individual detectors and move the array farther away from the target. This however increases the cors of the array.

\section{Conclusion}
We have simulated the efficiency to detect the full energy of light charged particles in CsI crystals using the HiRA10, LASSA and MUST2 telescopes. In order to correctly determine the efficiency of the experimental setup one has to carefully determine the efficiency loss due to multiple scattering as well as the nuclear reactions that occur in the detector. This effect is particularly important when using long CsI crystals ($>$6 cm) where the efficiency drops by more than 30\% (Fig.~\ref{fig:Eff_Range}). We evaluate the options available within the GEANT4 environment, find the more accurate options and evaluate their accuracy. It is interesting to note that the efficiency decreases almost exponentially as a function of the range and is rather similar for all the light charged particles. 

We find that multiple scattering decreases the efficiency significantly, especially near the boundary and for protons. Multiple scattering effects depend strongly on the geometry of a close-packed array. This geometry dependence indicates that efficiency losses should be an important consideration for the design of a close-packed array of CsI crystals. 

The best way to verify the efficiency results presented in the present paper would be to measure them experimentally using different mono-energetic charge particle beams. It would be interesting to perform such measurements for long CsI(Tl) crystals such as those for the HiRA10. Such measurements could provide both a direct efficiency measurement and a calibration of the energy vs. light output response of the CsI(Tl) crystals to these particles. In the case of protons, this could be achieved by measuring with protons at beam energies up to the full energy range of the crystal ($\approx$ 200 MeV). 

\section*{Acknowledgments}
This work is supported by the NSF under Grant No. PHY-1565546. The authors would like to thank the HiRA group at NSCL for the helpful discussion and suggestion.


\appendix
\section{}
\label{appendixA}
The transport of an incident flux of a particular species of charged particles through a CsI scintillator satisfies a Boltzmann equation that is modeled in GEANT4 by Monte Carlo techniques. Coulomb and nuclear elastic scattering deflect these particles and degrade their energies, while conserving flux. Nuclear reactions, however, typically change the charges, masses and energies of particles significantly. This often leads to the mis-identification of the particle, removing the flux of particles whose energies can be measured by stopping them in the detector.

We denote $P(X)$ as the probability the incident particle passes through a thickness of the detector $X$ without reaction:
\begin{equation}
P(X)=e^{-\lambda(X)},
\label{eq:prob}
\end{equation}
where
\begin{equation}
\lambda(X)=\int_0^X \rho\sigma(E(x))dx.
\label{eq:lambda}
\end{equation}
Here $\rho$ is the density of the material in atoms/cm$^3$, $x$ the distance that the particle has penetrated the material and $\sigma(E)$ is the energy-dependent reaction cross section for particles species reacting with atoms that compose the scintillator material. In the limit of a constant cross section, $P(X)$ becomes
\begin{equation}
P_{\sigma}(X) = e^{-\rho\sigma X}
\end{equation}
and the total efficiency including multiple scattering and reactions $\epsilon_{tot} (E)$ can be written as
\begin{equation}
\epsilon_{tot} (E) = \epsilon_{mult}(E) \times P_{\sigma}(R(E)),
\end{equation}
where $R(E)$ is the range of the particle with energy $E$ and $\epsilon_{mult}$ takes the efficiency loss due to multiple scattering out of the detector into account.\\
Alternatively, one can compute changes in $P(x)$ and in the total efficiency as the result of a series of probability losses $\Delta P = -\rho \sigma(E(x))P(x)\Delta x$ that occur within a section of the trajectory of length $\Delta x$ using the output of the Geant4 simulation. Using this approach, we simulated the different particles in Geant4 with only the Coulomb multiple scattering through the {``option4''} as explained in section 2.1. Since the reaction cross section does not change trajectories of non-reacted particles in the simulation, the reaction loss $\Delta P$ is then obtained from an analysis of the multiple scattering simulation. After the $i^{th}$ step in the analysis of an event in the simulation we get the energy deposited by the particle in the CsI crystal $\Delta E(E(x), \Delta x)_{loss,i}$ and the mean energy $E(x)$ of the particle during the step. We then calculate the value of the fractional probability $\Delta P$ using the reaction cross section at that energy and survival of the particle in this event from the following three steps
\begin{itemize}
\item We calculate a random number $p$ between 0 and 1,
\item if $p>| \Delta P |$, we assume there is no nuclear reaction and the event is kept,
\item if $p<| \Delta P |$, we assume that a nuclear reaction occurs and the event is terminated.
\end{itemize}

If no nuclear reaction has occurred in the thickness $\Delta x$ traversed during the step, we continue the process using the new energy of the particle $E_{i+1}=E_{i}-\Delta E(E(x), \Delta x)_{loss,i}$ until it stopes in the scintillator or scatters out of the crystal. If a reaction occurs in the i$^{th}$ step, then this reduces the number of properly detected particles by one event. In our work we used this Monte Carlo approach to calculate the efficiency for deuterons using the {``d-fit''} parameterization employing a step size corresponding to $\Delta x=$ 100 $\mu$m.

\begin{figure}[h!]
\begin{center}
\includegraphics[width=14.cm]{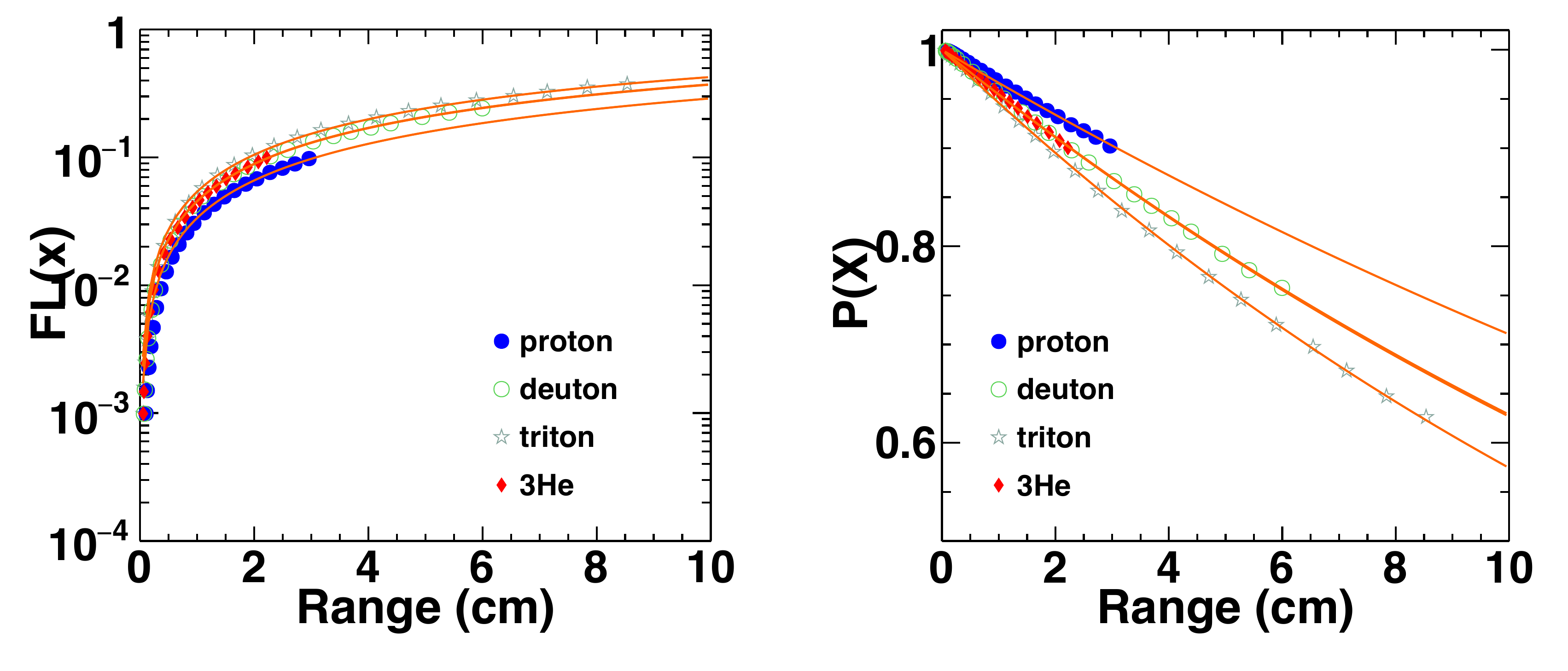}
\caption{\label{fig:FractionLoss} (a) Fraction loss in a CsI for proton (red), deuteron (yellow) and triton (blue) particule as a function of the range in the material \cite{cha95}. The lines correspond to the fit of the points assuming a constant cross section. (b) Associated probability $P(X)$ using this parameterization.}
\end{center}
\end{figure}

In Ref.~\cite{cha95}, Charity et. al calculated the {``fractional loss''} for various particles in CsI from nuclear reactions defined by $FL(E)=1-P(R(E))$. Their fractional losses are displayed by the points in Fig.~\ref{fig:FractionLoss}a) as a function of the range $R$. To obtain the lines in Fig.~\ref{fig:FractionLoss}a), we fitted their values of $FL(E)$ by assuming constant reaction cross sections obtaining values for these cross sections, which are given in Tab.~\ref{tab:charity}; these values are comparable to those given in Perey and Perey \cite{per76} for target nuclei in Cs or I mass region. The fractional loss in Fig.~\ref{fig:FractionLoss}a) corresponds to the efficiency displayed in Fig.~\ref{fig:FractionLoss}b); where the efficiency losses due to multiple scattering are neglected as in the calculations of Ref.~\cite{cha95}. The efficiencies, labelled as {``Charity''} in Fig.~\ref{fig:Eff_H}, however include the efficiency losses due to multiple scattering in addition to those cause by reactions.

\begin{table}[h!]
\begin{center}
\begin{tabular}{c | c }      
\hline \hline     
 Particle   & $\sigma$ (barn)\\ \hline 
$p$		& 1.6	\\
$d$		& 2.2\\
$t$		& 2.7	\\
$^{3}$He  & 2.2	\\
$\alpha$  	& 2.3	\\
\hline \hline 
\end{tabular}
\caption{\label{tab:charity} Fit of the fractional loss (Fig. \ref{fig:FractionLoss}) due to nuclear reaction in CsI material \cite{cha95}. The nuclear cross section for a given particle is also reported}
\end{center}
\end{table}



\newpage
\begin{thebibliography}{99}
\bibitem{lassa} A. Wagner {\it et al.} {\it Nucl. Instr. and Meth. A} {\bf 456} 290-299 (2001)
\bibitem{lassa1} B. Davin {\it et al.} {\it Nucl. Instr. and Meth. A} {\bf 473} 302-318 (2001)
\bibitem{must} Y. Blumenfeld {\it et al.} {\it Nucl. Instr. and Meth. A} {\bf 421} 471-491 (1999)
\bibitem{must2} E. Pollaco {\it et al.} {\it Eur. Phys. J. A.} {\bf 25} 287-288 (2005)
\bibitem{hira07} M. S. Wallace {\it et al.} {\it Nucl. Instr. and Meth. A} {\bf 583} 302-312 (2007)
\bibitem{tsa01} M. B. Tsang {\it et al.} {\it Phys. Rev. C} {\bf 64} 054615 (2001)
\bibitem{tsa09} M. B. Tsang {\it et al.} {\it Phys. Rev. Lett.} {\bf 102} 122701 (2009)
\bibitem{cou16} D. D. Coupland {\it et al.} {\it Phys. Rev. C} {\bf 94} 011601(R) (2016)
\bibitem{zha14} Y. Zhang {\it et al.} {\it Phys. Lett. B} {\bf 732} 186-190 (2014)
\bibitem{cha95} R. J. Charity {\it et al.} {\it Phys. Rev. C} {\bf 52} 3126 (1995)
\bibitem{wag01} A. Wagner {\it et al.} {\it Nucl. Instr. and Meth. A} {\bf 456} 290-299 (2001)
\bibitem{goe04} M.-J. van Goethem {\it et al.} {\it Nucl. Instr. and Meth. A} {\bf 526} 455-476 (2004)
\bibitem{usu16} Z. Usubov arXiv:1604.00827 (2016)
\bibitem{npt16} A. Matta {\it et al.} {\it Journal. Phys. G: Nucl. Part. Phys} {\bf 43} 045113 (2016)
\bibitem{ROOT} R. Brun and F. Rademakers {\it Nucl. Instrum. Methods Phys. Res. A} {\bf 389} 6-81 (1997)
\bibitem{Geant4} S. Agostinelli {\it et al.} {\it Nucl. Instrum. Methods Phys. Res. A} {\bf 506} 250-303 (2003)
\bibitem{geant4_doc} http://geant4.web.cern.ch/geant4/UserDocumentation/UsersGuides/\\
PhysicsReferenceManual/fo/PhysicsReferenceManual.pdf
\bibitem{dw4} P. D. Kunz, computer code DWUCK4, University of Colorado (unpublised)
\bibitem{ch89} R. L. Varner {\it et al.} {\it Physics Report} {\bf 201} 57-119 (1991)
\bibitem{shen} Wen-qing Shen {\it et al.} {\it Nuclear Physics} {\bf A491} 130-1146 (1989)
\bibitem{tripathi} R. K. Tripathi {\it et al.} {\it NASA Technical Paper} TP-1999-209726 (1999)
\bibitem{glauber} V. M. Grichine {\it Eur. Phys. J. C} {\bf 62} 399-404 (2009)
\bibitem{devries} R. M. DeVries and J. C. Peng  {\it Physical Review C} {\bf 22} 1055 (1980)
\bibitem{incl} P. Kaitaniemi {\it et al.} {\it Progress in NUCLEAR SCIENCE and TECHNOLOGY} {\bf 2} 788-793 (2011)
\bibitem{carlson} R. F. Carlson {\it Atomic Data And Nuclear Data Tables} {\bf 63} 93-116 (1996)
\bibitem{madani} J. H. Madani {\it Nuclear Physics A} {\bf 839} 42-50 (2010)
\bibitem{pen81} J. C. Peng {\it Physics Letters} {\bf 98B} 244-247 (1981)
\bibitem{hon14} Jun Hong and P. Danielewicz {\it Phys. Rev. C} {\bf 90} 024605 (2014)
\bibitem{per76} C.M. Perey and F.G. Perey {\it At. Data Nucl. Data Tables} {\bf 17} 1 (1976)
\bibitem{ter73} Y. Terrien {\it Nuclear Physics A} {\bf 199} 65-80 (1973)
\end{thebibliography}
\end{document}